# Why do investors buy shares of actively managed equity mutual funds? Considering the Correct Reference Portfolio from an Uninformed Investor's Perspective[1, 2]

Radu **BURLACU**[3]
*Univ. Grenoble Alpes, Grenoble INP\*, CERAG, 38000 Grenoble, France*
*\*Institute of Engineering and Management, Univ. Grenoble Alpes*
*Radu.burlacu@univ-grenoble-alpes.fr*

Patrice **FONTAINE**
*CNRS EUROFIDAI, 38000 Grenoble, France*
*Léonard de Vinci Pôle Universitaire, Research Center, 92000 Paris La Défense, France*
*Patrice.fontaine@eurofidai.org*

Sonia **JIMENEZ-GARCÈS**
*Univ. Grenoble Alpes, Grenoble INP\*, CERAG, 38000 Grenoble, France*
*\*Institute of Engineering and Management, Univ. Grenoble Alpes*
*Sonia.jimenez-garces@univ-grenoble-alpes.fr*

---

[1] We would like to express our sincere thanks to the anonymous referees and editors for the considerable time they spent reading our paper. Their valuable comments and suggestions greatly improved our article.
[2] This paper has received financial support from the Dauphine Chair in Asset Management, an initiative of Amundi and the University Paris-Dauphine, under the aegis of the Dauphine Foundation.
[3] Corresponding author. E-mail address: radu.burlacu@univ-grenoble-alpes.fr.

# ABSTRACT

We use the Grossman & Stiglitz (1980) framework to build a reference portfolio for uninformed investors and employ this portfolio to assess the performance of actively managed equity mutual funds. We propose an empirical methodology to construct this reference portfolio using the information on prices and supply. We show that mutual funds provide, on average, an *insignificant* alpha of 23 basis points per year when considering this portfolio as a reference. With the stock market index as a proxy for the market portfolio, the average fund alpha is negative and highly significant, −128 basis points per year. The results are robust when considering various subsets of funds based on their characteristics and their degree of selectivity. In line with rational expectations equilibrium models considering asymmetrically informed investors and partially revealing equilibrium prices, our study supports that active management adds value for uniformed investors.

# RESUME

Nous utilisons le modèle de Grossman & Stiglitz (1980) pour construire un portefeuille de référence pour les investisseurs non informés et nous utilisons ce portefeuille pour évaluer la performance des fonds communs de placement en actions, gérés activement. Nous proposons une méthodologie empirique pour construire ce portefeuille de référence en utilisant des informations sur les prix des actions et l'offre en titres. Nous montrons alors que les fonds communs de placement fournissent, en moyenne, un alpha *non significatif* de 23 points de base par an lorsque l'on considère ce portefeuille comme référence. Avec l'indice du marché comme proxy du portefeuille de marché, le alpha moyen des fonds est négatif et hautement significatif (-128 points de base par an). Les résultats sont identiques lorsque l'on considère différents sous-ensembles de fonds en fonction de leurs caractéristiques et de leur degré de sélectivité. En accord avec les modèles d'équilibre à anticipations rationnelles considérant des investisseurs différemment informés et des prix d'équilibre partiellement révélateurs, nos résultats indiquent que la gestion active des fonds mutuels est performante du point des investisseurs non informés.

The mutual fund industry has shown rapid growth over the last decades, and this growth is particularly spectacular for actively managed equity mutual funds. Using data from the CRSP Survivor-Bias-Free US Mutual Fund from 1968 to 2020, we document a 12.67% average annual growth rate of total net assets for US-domestic actively managed equity mutual funds. Whether active mutual fund management adds value for investors is of considerable interest to academics and practitioners.

Many academic studies analyze the performance of actively managed equity mutual funds and find pessimistic results[4]. In most empirical specifications, these studies use the market index as the main reference portfolio for measuring mutual funds' performance. Funds appear to deliver, on average, a negative and significant market-risk-adjusted return (alpha). The conventional view is that the performance of active mutual fund management is negative. An important and long-debated question remains unanswered: *Why do investors continue to buy shares of actively managed mutual funds knowing that their performance is negative?*

Our study answers this question by mobilizing rational expectations equilibrium (REE) models[5] considering asymmetrically informed investors and partially revealing assets' prices. These models provide a relevant theoretical framework for analyzing mutual fund performance. In particular, the Grossman and Stiglitz (1980) model, hereafter GS, considers the existence of two groups of investors on the market: informed investors and uninformed ones. Mutual fund managers can be thought of as representing the group of informed investors in the GS model while their clients can be considered as the group of uninformed investors.

In this paper, we claim that the market portfolio is *not* the correct reference portfolio for assessing portfolio performance, neither from the perspective of informed investors (mutual fund managers) nor from the perspective of uninformed ones (mutual fund clients). In a market with informational asymmetries between groups of investors, each group has its own efficient frontier, asset-pricing model, and optimal portfolio conditional on the available information. This is a well-known result of REE models. In particular, the optimal portfolio of uninformed inves-tors under-weighs (respectively overweighs) assets with high (respectively low) degrees of information asymmetry. The performance of any asset, and in particular that of mutual funds, must be measured relative to this reference portfolio if one takes the perspective of an uninformed investor. Using the stock market index as a proxy for the market portfolio in empirical studies would translate into a biased performance (alpha). In particular, this may explain the pessimistic results about mutual funds' performance obtained in the literature.

---

[4] We can cite, among others, Jensen (1968), Henriksson (1984), Chang & Lewellen (1984), Grinblatt & Titman (1993), Malkiel (1995), Gruber (1996), Grinblatt *et al.* (1995), Carhart (1997), Wermers (1997), as well as Chen *et al.* (2000).

[5] Grossman (1976), Grossman & Stiglitz (1980) and Hellwig (1980) develop these types of models in one-risky-asset frameworks. Admati (1985), O'Hara (2003), and Easley & O'Hara (2004), among others, extend these models in multi-risky-asset frameworks.

Our objective in this paper is to use the reference portfolio of uninformed investors to assess the performance of actively managed mutual funds from the perspective of these investors, considered clients. As far as we know, our paper is the first that uses such a portfolio derived from REE models to assess mutual fund performance. We designate this portfolio "UIRP" for "Uninformed Investor's Reference Portfolio".

Specifically, our paper brings two major contributions to the literature. The first is theoretical; we build on existing multi-asset generalizations of the GS model (Jimenez-Garcès, 2004) to build "UIRP". To achieve this goal, we proceed in three steps. First, we explicitly derive the asset-pricing model from the uninformed investors' perspective. From their perspective, the expected return of an asset in excess of the risk-free rate equals the asset's beta with respect to UIRP multiplied by the expected return of this reference portfolio in excess of the risk-free rate (UIRP risk premium). The beta with respect to the market portfolio and the market risk premium, which are derived from the classical CAPM, is irrelevant for pricing assets from the uninformed investor's perspective; the relevant reference portfolio is UIRP. The second step is to determine UIRP based on *unobservable* REE parameters. We analyze particular cases and perform simulation analyses showing that UIRP under-weighs (respectively overweighs) assets with high (respectively low) levels of information asymmetry. The third step consists in determining UIRP based on *observable* variables. For measuring the degree of information asymmetry, we use the "Proxy E(r)" measure proposed by Burlacu *et al.* (2012). This measure obtains by projecting individual common stock returns on their prices and the prices of relevant industry portfolios. Based on a simulation analysis, we prove the relevancy of this measure for building UIRP.

The second contribution of this paper is empirical. It consists in analyzing the performance of domestic U.S. actively managed equity mutual funds by using UIRP as a reference portfolio. To achieve this goal, we proceed in three steps. First, we use CRSP common stock data on the U.S. market to compute "Proxy E(r)" for the 53 years from January 1968 to December 2020.

Second, we construct UIRP by following the guidelines from the theory. This portfolio is not investable as it does not use only *ex-ante* publicly available information (beginning-of-period prices) but also information on supply (stock capitalization), which does not represent *ex-ante* observable information for uninformed investors in the theoretical framework considered here. Rather than giving a tool for portfolio management to these investors, our aim in building this portfolio is to provide a tool allowing them to assess the performance of actively managed portfolios on an *ex-post* basis. Third, we analyze the performance of 26,407 domestic U.S. actively managed equity mutual funds existing from January 1968 to December 2020 by using CRSP survivorship-free mutual fund data. The mutual fund's performance is estimated with UIRP and then is compared to the results obtained using alternative portfolios such as the market index (Jensen, 1968) or other widely used factor-mimicking portfolios (Carhart, 1997).

Our empirical analysis is deployed for mutual funds overall and categories of mutual funds based on their characteristics such as expenses, turnover, total net assets, or degree of selectivity. Using UIRP, we find that actively managed equity mutual funds offer their clients a non-significant alpha of 23 basis points per year. Our results are robust when considering specific categories of mutual funds based on their characteristics. In addition, we do not find significant performance for groups of mutual funds based on their degree of active management (selectivity). Using the market index as a reference portfolio provides a different picture but closer to that offered by the existing literature. The average actively managed equity fund provides a negative and highly significant alpha of -128 basis points per year. We also find negative performance for categories of mutual funds based on their characteristics and degree of selectivity.

Overall, our results are in line with the theory and support the value of active management. Our study joins a branch of the literature that provides a more optimistic view about mutual fund active management. Some studies argue that the poor performance of actively managed funds, as put forward by the previous literature, can be explained by the services that mutual funds bring to investors. For example, Edelen (1999) considers the costs associated with the liquidity service provided by mutual funds. By controlling for such costs, the abnormal returns obtained by mutual fund clients become insignificant. Other empirical studies measure fund performance at the portfolio's stockholdings level. Wermers (2000) shows that the portfolio of common stocks held by actively managed equity mutual funds outperforms the stock market. The over-performance obtained at the fund's stockholding level just compensates for mutual fund ex-penses and transaction costs. Wermers (2000) concludes that this result is in line with the predictions of the GS model. Our study achieves the same conclusion as in Wermers (2000), but our approach differs. We argue that the poor performance put forward by the literature on actively managed funds comes from the fact that this literature uses a reference portfolio that is unsuitable if one takes the perspective of REE models with asymmetrically informed investors.

The paper's structure is as follows. Section 1 derives the uninformed investor's asset pricing model and his optimal portfolio based on a generalization of the GS model in a multi-asset framework. In section 2, we propose an empirical methodology to construct UIRP. Section 3 analyzes mutual fund performance using UIRP as a reference portfolio together with other widely used reference portfolios. The last section concludes and offers some avenues for future research.

# 1 BUILDING UIRP: THEORETICAL FRAMEWORK

This section starts by justifying our approach. Then, it briefly describes the multi-asset version of the GS model developed by Jimenez-Garcès (2004) by focusing on the two main issues of interest: *What is the asset-pricing model, and what is the reference portfolio for an uninformed investor in this model?* This section ends by proposing a methodology to build UIRP using observable variables. Subsequently, we will use UIRP to measure fund performance from the perspective of uninformed investors, considered as clients buying shares of funds. This represents the main objective of this paper.

## 1.1 JUSTIFICATION FOR THE USE OF REE MODELS

Consider an "uninformed" investor who envisages the possibility of buying shares of actively managed mutual funds. What is the asset-pricing model from the perspective of this investor? What is his optimal portfolio? The traditional Sharpe–Lintner–Mossin capital asset pricing model (CAPM) assumes that all investors possess homogeneous information about the future returns of market-traded securities. It predicts that the expected return of any asset depends only on its systematic (market) risk and that all investors hold a combination of the risk-free asset and the market portfolio. In this framework, the optimal portfolio strategy for any investor is to buy-and-hold the market portfolio.

The empirical literature does not provide strong support for the frictionless CAPM's predictions, neither in terms of asset pricing nor in terms of portfolio management. Measures of systematic risk do not explain common stocks' expected returns well, and the popularity of active portfolio management represents a departure from the predicted passive buy-and-hold strategy. In contrast, theories of financial markets considering asymmetrically informed investors seem well supported by existing empirical studies. Measures of information risk put forward by this literature seem to perform well in explaining assets' returns. For example, the "Probability of Information-Based Trading" (PIN) introduced by Easley *et al.* (2002), a measure of "private information" based on a microstructure model, is statistically and economically significant in explaining U.S. common stock returns. The predictions of REE models are also supported in terms of portfolio management (e. g., Biais *et al.*, 2010).

When investors are differently informed, analyzing the performance of mutual funds needs to be considered in a more realistic setting than the traditional CAPM. Active mutual fund managers spend time, expertise, and resources on acquiring and processing information. Their clients, who do not have the managers' ability to process information, should be considered uninformed investors. Based on their expertise, fund managers should be able to make better portfolio allocation decisions than uninformed investors. REE models predict that, at equilibrium, the extra performance obtained by mutual funds just compensates the information-acquisition costs incurred.

The partition of the investors' universe in these two groups (informed and uninformed investors) justifies why we use REE models with asymmetrically informed investors, specifically the GS model, for our research issue. Our objective is to mobilize the GS model to build a reference portfolio for assessing fund performance from the point of view of uninformed investors explicitly[6].

## 1.2 THE SETTING

GS consider only one risky asset. We use the multi-asset version of this model as proposed by Jimenez-Garcès (2004), who presents an exact multi-asset replication of the original GS model[7]. The model is static and considers *two* types of investors, informed and uninformed, who are rational (they maximize the utility of their final wealth conditionally on the available information set). Agents invest at $t = 0$ and consume at $t = 1$. There are $n$ risky securities and one risk-free asset. For simplicity, the risk-free rate of return is set equal to zero. The percentage of informed investors is denoted by $\lambda$. The securities' payoff $\tilde{P}_1$ at $t = 1$ writes:

$$\tilde{P}_1 = \tilde{\theta} + \tilde{\varepsilon} \tag{1}$$

where $\tilde{\theta}$ is the payoff's component known only by informed investors, designated "information", and $\tilde{\varepsilon}$ is the component unknown to all investors, designated "residual uncertainty". The expected value of the residual uncertainty equals zero. The vectors $\tilde{\theta}$ and $\tilde{\varepsilon}$ are jointly multivariate normal and independent, and their variance-covariance matrices are denoted by $T$ and $U$, respectively[8]. The information set of the informed investor resides in the observation of $\tilde{\theta}$ and that of the equilibrium price at $t = 0$. The information set of the uninformed investor consists only of the observation of the equilibrium price at $t = 0$.

---

[6] Other models consider that the information is dispersed across investors rather than detained by just one group of them (Hellwig, 1980; Admati, 1985; Fontaine *et al.*, 2018). Other theories rationalizing optimal portfolios that deviate from the market portfolio include those considering non-standard preferences, such as the prospect theory, or theories considering disappointment aversion (e.g., Dahlquist *et al.*, 2016; Ang *et al.*, 2005).

[7] Kodres & Pritsker (2002) also propose a multi-asset generalization of the GS model. Their model departs from GS by considering that noise comes from the presence of liquidity traders, not from supply uncertainty.

As is common in *REE* models, per capita asset supply is a random vector $\tilde{z}$. This vector is independent of $\tilde{\theta}$ and $\tilde{\varepsilon}$ and joins the normal distribution of the other variables in the model. The assumption that $\tilde{z}$ is unknown by investors is necessary for making equilibrium prices partially (not perfectly) revealing[9]. The variance-covariance matrix of the supply is designated by $Z$. Matrices $U, T,$ and $Z$ are regular, which is a necessary condition for ensuring the existence and the uniqueness of the equilibrium price. In addition, it is assumed that investors have constant absolute risk aversion, implying that their demand is independent of their initial wealth. The risk aversion coefficient $a$ is identical for all investors, $a > 0$.

Under the above conditions, following the same line of reasoning as in GS, there is a unique closed-form solution for the equilibrium price $\tilde{P}^0$ (at $t = 0$) within the class of linear functions of $\tilde{\theta}$ and $\tilde{z}$ of the form:

$$\tilde{P}^0 = A_0 + A_1[\tilde{\theta} - \alpha U \tilde{z}] \tag{2}$$

where $\alpha = a/\lambda$,

$$A_0 = (I_n - A_1)E(\tilde{\theta}) + [\alpha A_1 U - a V_m]E(\tilde{z})$$

---

[8] Our theoretical framework comes with all the caveats associated with considering that asset prices are normally distributed, which in particular implies that they can be negative, and with considering constant absolute risk aversion. This may generate anomalies, already studied in Admati (1985).

[9] See the « Grossman and Stiglitz paradox » (Grossman, 1976) for a discussion on this aspect.

$$A_1 = (T + U - V_m)T^{-1}$$

$$V_m = [\lambda U^{-1} + (1-\lambda)(N+U)^{-1}]^{-1}$$

$$N = Var(\tilde{\theta}|\tilde{P}^0) = (T^{-1} + \alpha^{-2}U^{-1}Z^{-1}U^{-1})^{-1}$$

Matrix $V_m$ is the average variance-covariance matrix of assets' payoffs. The latter is the harmonic mean of the informed and uninformed investors' variance-covariance matrices, $Var(\tilde{P}^1|\tilde{\theta}) = U$ and $Var(\tilde{P}^1|\tilde{P}^0) = N + U$, respectively.

Eq. (2) shows that the equilibrium price $\tilde{P}^0$ is a linear function of two random variables: information ($\tilde{\theta}$) and supply ($\tilde{z}$). Because supply is unknown, uninformed investors are unable to infer the realization of $\tilde{\theta}$ by observing assets prices. Thus, some information remains private, generating informational asymmetries between investors. The degree of information asymmetry is represented by matrix $N = Var(\tilde{\theta}|\tilde{P}^0)$. A higher variance of information conditional on prices means a higher degree of information asymmetry between informed and uninformed investors.

## 1.3 THE UNINFORMED INVESTOR'S ASSET PRICING MODEL

Uninformed investors maximize their utility conditional on public information (i.e., the information provided by equilibrium prices). The number of risky securities in their optimal portfolio writes:

$$\tilde{X}_{UI} = a^{-1}Var_{UI}^{-1}(\tilde{P}^1 - \tilde{P}^0) \times E_{UI}(\tilde{P}^1 - \tilde{P}^0) \tag{3}$$

The subscript $UI$ indicates that the variables are conditional on the information detained by uninformed investors. After multiplying the left-hand side of this equation by $Var_{UI}(\tilde{P}^1 - \tilde{P}^0)$, it follows

$$E_{UI}(\tilde{P}^1 - \tilde{P}^0) = aCov_{UI}(\tilde{P}^1 - \tilde{P}^0; \tilde{P}^1_{opt,UI} - \tilde{P}^0_{opt,UI}) \tag{4}$$

where $\tilde{P}^1_{opt,UI}$ is the payoff of uninformed investors' optimal portfolio and $\tilde{P}^0_{opt,UI}$ is the current value of this portfolio. The expected return from these investors' perspective for a given asset $i$ equals the covariance between the return of this asset and the return of their optimal portfolio.

In Eq. (4), returns are expressed as price differences. A more familiar specification involves returns expressed as the ratio between price differences and beginning-of-period prices. After classical manipulations, we obtain the following asset-pricing model for an asset $i$

$$E_{UI}(\tilde{r}_i) = \beta_{UI,i} E_{UI}(\tilde{r}_{opt,UI}) \tag{5}$$

where $\beta_{UI,i} = Cov_{UI}(\tilde{r}_i; \tilde{r}_{opt,UI})/Var_{UI}(\tilde{r}_{opt,UI})$ and $\tilde{r}_{opt,UI}$ is the return of the uninformed investor's optimal portfolio, which we have previously designated UIRP. From the perspective of these investors, the expected return for an asset $i$ depends linearly on its beta relatively to UIRP and the expected return of UIRP. It is important to emphasize that the risk premium required by uninformed investors to hold an asset (particularly a share of a mutual fund) is determined by UIRP, *not* by the market portfolio. As shown in Eq. (5), in a GS world, the alpha obtained by regressing (net of expenses) fund returns on the returns of the price-contingent portfolio UIRP has to be indistinguishable from zero. This important result justifies our paper's approach to assessing the performance of mutual funds.

Another aspect that deserves to be mentioned is that, in the GS framework, the market portfolio performs better than UIRP. The market portfolio is indeed the weighted average of two portfolios, UIRP, and the optimal portfolio of informed investors (which performs better than UIRP because these investors have an informational advantage), the weights being the proportions of the two categories of investors on the market. It follows immediately that, in the GS world, the alpha obtained by regressing (net of expenses) fund returns on the market portfolio returns would be *negative*. Existing studies largely corroborate this prediction, as they find negative market-risk-adjusted returns for actively managed mutual funds. Yet, this does not mean that the mutual fund is poorly managed; it just comes from the fact that the reference portfolio utilized in assessing performance is not the correct one.

## 1.4 THE UNINFORMED INVESTOR'S OPTIMAL PORTFOLIO

We now analyze the structure of the UIRP price-contingent portfolio. To simplify the analysis, we focus on the unconditional expectation of the portfolio's composition[10]. After some simple mathematical manipulations, it follows

$$E(\tilde{X}_{UI}) = (I + \lambda U^{-1} N)^{-1} * E(\tilde{z}) \tag{6}$$

The optimal portfolio's composition depends on the degree of information asymmetry between investors, assessed by the matrix $N$, the residual uncertainty, $U$, and the expected value of the number of assets per capita, $E(\tilde{z})$.

---

[10] Considering conditional expectations makes it difficult to analyze asset prices and optimal portfolios. Furthermore, anomalies may arise, which have been studied extensively in the literature (Admati, 1985; Jimenez-Garcès, 2004).

Eq. (6) suggests that the number of securities in the price-contingent portfolio is negatively related to their degree of information asymmetry, as assessed by the matrix $N$. Nevertheless, the equation is difficult to interpret because the portfolio's composition is expressed with matrices. One level of analysis is to consider that all model's exogenous matrices ($T$, $U$ and $Z$) are diagonal, which means that securities are not correlated. In this case, Eq. (6) implies that the number of shares of asset $i$ in the portfolio equals $(1 + \lambda u_i^{-1} n_i)^{-1} * E(\tilde{z}_i)$, where $u_i$ is the residual variance of asset $i$ and $n_i$ is the variance of $\tilde{\theta}_i$ conditional on prices. It follows immediately that a higher information asymmetry for asset $i$ implies a lower weight for this stock in the uninformed investor's optimal portfolio.

Building on REE models, Burlacu *et al.* (2012) propose a measure that is positively correlated to the degree of information asymmetry of a stock. This measure, known as "$Proxy\ E(r)$", is the $R^2$ from the regression of stock returns on the beginning-of-period prices. "$Proxy\ E(r)$" can be easily determined using only stock prices and proves to be relevant on both theoretical and empirical grounds as a measure of information asymmetry. Simulation analyses performed by Jimenez-Garcès (2004), who uses the GS setting, and Burlacu *et al.* (2012), who use Admati's (1985) setting, show that this measure captures information asymmetry effects. Stocks with high (low) information asymmetry levels have high (low) $R^2$ and exhibit high (low) expected returns. This result is robust when considering various levels of correlation between assets' information, residual uncertainty, or supply uncertainty. On empirical grounds, "$Proxy\ E(r)$" proves to be a statistically significant predictor of expected returns and is robust to size and book-to-market ratios, liquidities, and other measures of information asymmetry such as the probabilities of information-based trading[11].

---

[11] Biais *et al.* (2010) also propose a methodology that uses the information contained in prices to build the optimal portfolio from the perspective of uninformed investors. The methodology extracts the information contained in prices by projecting stock returns on prices for six portfolios constructed from a double sort of stocks based on the firm's size and its book-to-market value. The expected returns and variance–covariance matrix from this projection allow constructing the optimal portfolio of the uninformed investor. This approach implicitly considers that all stocks belonging to a given portfolio among the six portfolios have the same level of information asymmetry because these portfolios are, by construction, value-weighted. Our approach rather considers distinctly the level of information asymmetry for each individual stock, not for portfolios of stocks. For this reason, we believe that our approach allows constructing a reference portfolio that is more adequate for assessing the performance of mutual funds.

We use "$Proxy\ E(r)$" to construct UIRP. As Eq. (6) suggests, the weight of a stock in UIRP is negatively related to its level of information asymmetry. Therefore, $1 - $ "$Proxy\ E(r_i)$" for stock $i$ will be considered as the weight of this stock in UIRP. This weight can be considered as a measure of information precision; a higher precision (i.e., a lower degree of information asymmetry) implies a higher weight of the stock in the UIRP price-contingent portfolio.

## 1.5 SIMULATION ANALYSIS

We now check that the portfolio built using this information asymmetry measure behaves similarly to the price-contingent portfolio based on the unobservable REE model parameters as in Eq. (6). The analysis cannot be performed analytically because of the complexities involved by matrix computations. Instead, we will use a simulation analysis. For a given stock $i$, $1 - $ "$Proxy\ E(r)$" represents the return variance conditional on prices divided by the total (unconditional) return variance:

$$1 - \text{"}Proxy\ E(r_i)\text{"} = Var(\tilde{P}_i^1 - \tilde{P}_i^0 | \tilde{P}^0) / Var(\tilde{P}_i^1 - \tilde{P}_i^0) \tag{7}$$

where $\tilde{P}_i^1 - \tilde{P}_i^0$ represents the return of stock $i$ expressed as a price difference. The numerator and the denominator in the right-hand side of Eq. (7) are the $i$-th terms on the main diagonal of matrices $Var(\tilde{P}^1 - \tilde{P}^0 | \tilde{P}^0)$ and $Var(\tilde{P}^1 - \tilde{P}^0)$.

By replacing the formula of the equilibrium price $\tilde{P}^0$ from Eq. (2), we obtain:

$$Var(\tilde{P}^1 - \tilde{P}^0 | \tilde{P}^0) = N \tag{8}$$

$$Var(\tilde{P}^1 - \tilde{P}^0) = (I - A_1)T(I - A_1) + \alpha^2 A_1 UZUA_1 + U \tag{9}$$

The terms of the right-hand side of Eq. (7) obtain from multiplying the matrices in Eq. (8) and Eq. (9) at the left and at the right by a vector in which entries are all equal to zero except the $i$-th entry, which equals one.

Our simulation analysis considers a market with two risky stocks $i \in \{1, 2\}$. The goal is to compare, for various levels of information asymmetry, the structure of the portfolio based on unobservable REE parameters, as given by Eq. (6), and that of the portfolio assessed with observable variables. The latter uses $1 - $ "$Proxy\ E(r)$" as a weight for each individual stock. We study how information asymmetry impacts the ratio of the two stock holdings in the price-contingent portfolio determined with unobservable REE parameters, $\frac{E(\tilde{X}_{UI}^1)}{E(\tilde{X}_{UI}^2)}$, and the ratio of the two stock holdings in the portfolio determined with observable variables, $\frac{[1-\text{"Proxy }E(r_1)\text{"}]*E(\tilde{z}_1)}{[1-\text{"Proxy }E(r_2)\text{"}]*E(\tilde{z}_2)}$, where $E(\tilde{z}_i)$ is the expected supply for stock $i$. On a perfect market, each investor optimally holds the market portfolio; the ratio of the two stocks equals $\frac{E(\tilde{z}_1)}{E(\tilde{z}_2)}$ for each investor. On a market with informational asymmetries, the uninformed investor is expected to hold less of the stock with a higher degree of information asymmetry. This aspect is confirmed by our simulation analysis hereafter.

We consider a risk-aversion coefficient $a = 0.1$ and a proportion of informed (resp. uninformed) investors of 0.4 (resp. 0.6). These choices are arbitrary and have no fundamental consequences on our results. The information uncertainty is modeled by the matrix $T = \begin{pmatrix} 1 & t_{12} \\ t_{12} & 1 \end{pmatrix}$ and the residual uncertainty is modeled by the matrix $U = \begin{pmatrix} 1 & u_{12} \\ u_{12} & 1 \end{pmatrix}$. The parameters $t_{12}$ and $u_{12}$ vary across simulations. Stock supply uncertainty is modeled by the matrix $= \begin{pmatrix} z_{11} & z_{12} \\ z_{12} & 1 \end{pmatrix}$.

We vary the level of information asymmetry on stock 1 by changing $z_{11}$; the latter takes values from 1 to 10. A higher level of noise (supply uncertainty) for stock 1 implies a higher level of information asymmetry for this stock. The latter is assessed by $n_1 = Var(\tilde{\theta}_1|\tilde{P}^0)$, and therefore we expect a lower weight of stock 1 in the price-contingent portfolio based on unobservable REE parameters and in the portfolio based on the observable variable $1 - \text{"Proxy }E(r)\text{"}$.

Table I shows the results of our simulation analyses for four configurations. The first one considers that stocks are not correlated: $t_{12} = u_{12} = z_{12} = 0$. The other three configurations consider a correlation coefficient of 0.5 for each one of the three parameters, with the other two correlation coefficients being equal to zero. For $z_{11} = 1$, the level of information asymmetry is identical for the two stocks, and therefore their weight in the portfolio is the same.

The table shows that, whatever the setting, an increase in noise implies an increase in the level of information asymmetry of stock 1 relative to stock 2, measured by $\frac{n_{11}}{n_{22}}$, where $n_{11}$ and $n_{22}$ are the diagonal terms of the matrix $N$. The table also shows that the ratio of the two stocks' weights in the price-contingent portfolio assessed with unobservable REE parameters and the one assessed with the observable variable $1 - \text{"Proxy }E(r)\text{"}$ are coherent. The two ratios decrease in a similar way, confirming that a higher level of information asymmetry on stock 1 implies a lower holding of this stock in the price-contingent portfolio, for both specifications.

(Insert Table I about here)

In unexposed analyses, we have performed various other simulations. In some of them, we have changed the total level of information for asset 1, $t_{11} = Var(\tilde{\theta}_1)$, and have found similar results. As $t_{11}$ increases, the information asymmetry for stock 1 decreases relative to stock 2. The explanation is that the amount of total information increases for a given level of noise, making the price of stock 1 more informative than the price of stock 2. The proportion of stock 1 in the price-contingent portfolio increases, whatever the portfolio specification. Another set of unexposed simulations consists in varying the level of residual uncertainty for stock 1, $u_{11} = Var(\tilde{\epsilon}_1)$, which is the first entry of the residual uncertainty matrix $U$. Again, we find similar results. Residual uncertainty affects all investors in the market. A high residual uncertainty "dilutes" the information asymmetry problem and, therefore, "aligns" the portfolios of all investors. The information asymmetry problem becomes less important for uninformed investors, which results in a higher proportion of stock 1 holdings in their portfolio.

Overall, we conclude that the observable variable $1 - "Proxy\ E(r)"$ performs well in structuring a portfolio that is close to the one held by uninformed investors in the GS framework. In what follows, we describe the empirical methodology employed to build this portfolio.

## 2 BUILDING UIRP: EMPIRICAL METHODOLOGY

This section presents our empirical proxy for the degree of information asymmetry affecting a stock. We use this proxy to build our uninformed investor's reference portfolio UIRP. As this portfolio uses *ex-post* information (assets' supply, which is proxied by common stocks' capitalization), this portfolio is not investable. Indeed, an uninformed investor does not observe supply. However, these investors can use this portfolio to assess the performance of their shares of funds. The market portfolio would be irrelevant for this purpose.

### 2.1 THE EMPIRICAL PROXY FOR THE DEGREE OF INFORMATION ASYMMETRY

We follow the theory and construct the reference portfolio from an uninformed investor's perspective by under- (over-) weighing common stocks with high (low) information asymmetry. We first estimate $Proxy\ E(r)$, which is the coefficient of determination[12] $R^2$ of the regression

---
[12] In what follows, we will use interchangeably $Proxy\ E(r)$ and $R^2$.

of stock returns on prices. The regression's specification is, as in Burlacu *et al.* (2012),

$$r_{i,t} = \alpha_{i,0} + \beta_{i,1} P_{i,t-1} + \sum_{j=1}^{4} \beta_i^{SICj} P_{i,t-1}^{SICj} + \varepsilon_{i,t} \quad (10)$$

where $r_{i,t}$ is the daily return of common stock $i$ on day $t$, $P_{i,t-1}$ is the closing (adjusted and normalized) price of stock $i$ on day $t-1$, $P_{i,t-1}^{SICj}$ is the (adjusted and normalized) price of industry portfolios on day $t-1$, and $\varepsilon_{i,t}$ is the error term.

The empirical variables are determined exactly as in Burlacu *et al.* (2012). The normalized price $P_i$ of an individual stock $i$ is determined in two steps. First, the price is set to one the first day the stock appears in the CRSP U.S. Daily Stock Database and then increased or decreased by the daily stock return, which is directly extracted from this database. Second, the price thus obtained is normalized, that is, we divide it by the price of the market index. We normalize prices to mitigate stationarity issues[13]. The price of the market index is set to one the first date it appears in the database in January 1967, and then it is increased or decreased by the daily return of the CRSP value-weighted index.

The normalized prices of industry portfolios are obtained in the same way as for individual stocks. The normalized price of the four-digit industry portfolio, $P_{i,t}^{SIC4}$, is calculated using stocks having the same four-digit standard industrial classification (SIC) as stock $i$ and excludes this stock. This exclusion avoids spurious correlation between variables. The portfolio price is set to one the first day it appears in our data set and then increased or decreased by the daily weighted-average portfolio's return. The price of industry portfolios represents the value of a buy-and-hold strategy consisting of investing one dollar at the beginning of the estimation period. The dividends are reinvested each period. The price is then normalized, that is, we divide it by the (normalized) price of the stock market index. The normalized price of the second portfolio, $P_{i,t}^{SIC3}$, is calculated using stocks having the same three-digit SIC code as stock $i$ but excludes all stocks having the same four-digit SIC code (which implies that stock $i$ is also excluded). The same procedure applies to the two-digit and one-digit industry portfolios. For more information on the construction of these portfolios, the reader can refer to Burlacu *et al.* (2012).

---

[13] Biais *et al.* (2010) analyze stationarity issues for their regressions of stock returns on the prices of Fama & French (1993) size and B/M portfolios. By means of Monte-Carlo simulations, they show that these regressions are not affected by stationarity. In analyses not included here, we have also investigated stationarity issues for our regressions. Our tests of stationarity and co-integration corroborate the results found by Biais *et al.* (2010).

The data cover the period from January 1967 to December 2020. We estimate stocks' $R^2$ as the fit from the regression shown in Eq. (10) for all American-listed ordinary common stocks, which CUSIP ends with 10 or 11. $R^2$ is estimated each month $m$ for each individual common stock $i$ using past daily returns and (adjusted and normalized) prices from months $m - 12$ to $m - 1$. We eliminate observations with zero stock returns to mitigate problems related to non-synchronous or "thin" trading[14]. With this restriction, we require a minimum of 60 daily returns for the computation of $R^2$. We are thus able to calculate $R^2$ for each stock and each month during the analysis period. $R^2$ will be considered the degree of information asymmetry affecting the stock during that month.

Based on our empirical design, we compute $R^2$ for 24,596 ordinary common stocks on the period between January 1968 and December 2020, with an average of 4,084 stocks per month. The mean $R^2$, which gauges the mean level of information asymmetry in the market, equals 6.42%, with a standard deviation of 4.93%. Its median value is 5.24%, and the intra-quartile range is [3.35%, 8.04%].

## 2.2 CONSTRUCTION OF UIRP

Since $1 - R^2$ is inversely correlated to the degree of information asymmetry as measured by $n_i$, it will be used to quantify the weight of stock $i$ in UIRP. The return of this portfolio is a modified version of the market portfolio in which the weight of a given common stock is related to its capitalization and its degree of informational asymmetry, as indicated by the theory:

$$r_{UIRP} = \frac{\sum_{i=1}^{n}(1 - R_i^2)K_i r_i}{\sum_{i=1}^{n}(1 - R_i^2)K_i} \tag{11}$$

In this equation, $r_{UIRP}$ represents the return of the uninformed investor's reference portfolio, $K_i$ is the capitalization of common stock $i$, and $R_i^2$ obtains from the regression of stock $i$'s returns on prices as explained in the previous subsection.

It is important to note that UIRP is not investable from the perspective of uninformed investors because it uses *ex-post* information. The weights used in equation (11) are indeed based on the market capitalization of common stocks. In the GS framework, uninformed investors do not observe market supply and, therefore, cannot construct this portfolio on an *ex-ante* basis. However, as affirmed previously, our objective is not to build an investable portfolio for uninformed investors. Instead, we aim to construct a reference portfolio allowing these investors to check *ex-post* that the return generated by an actively managed equity mutual funds portfolio is sufficient to compensate for the risk as perceived by uninformed investors. *Ex-post*, these investors obtain the information on market supply and are therefore able to construct this portfolio by following the methodology described in this paper. Of course, this portfolio is more difficult to build than the market portfolio, as it uses weights based on the stock's capitalization and its degree of informational asymmetry. However, UIRP is more relevant to consider as a reference portfolio because it is closer to the uninformed investor's optimal portfolio as predicted by the theory. The market portfolio is irrelevant as a reference portfolio in the presence of asymmetry information.

---

[14] We check the influence of non-synchronous trading by restricting our analysis to securities having 10 minimum trading days per month or by eliminating stocks with prices lower than $1. Imposing this restriction does not change qualitatively our conclusions.

Overall, UIRP is constructed using the universe of stocks belonging to the CRSP value-weighted stock index but they are weighted differently in comparison to the CRSP index. However, the lack of sufficient data implies that we are not able to compute the $R^2$ for all NYSE, AMEX, and NASDAQ-listed common stocks. We, therefore, construct UIRP in the following way. First, we build the portfolio as in Eq. (11) with NYSE, AMEX, and NASDAQ-listed common stocks for which $R^2$ can be determined. Then, we form a capitalization-averaged portfolio of the remaining common stocks listed on these markets, for which the data is not sufficient to determine $R^2$. UIRP is the weighted average of the two portfolios, the weights being their capitalizations. Our results do not fundamentally change if we consider the portfolio containing only stocks with available $R^2$. The stocks for which we are not able to determine the $R^2$ concern small firms, which weigh little in comparison to other stocks.

To assess the deviation between UIRP and the CRSP stock market index, we calculate the difference between the amount invested in the market index and the amount invested in UIRP, $\sum_{i=1}^{n} K_i - \sum_{i=1}^{n}(1 - R_i^2)K_i = \sum_{i=1}^{n} R_i^2 K_i$, and then we express it as a percentage of the total market capitalization, $\sum_{i=1}^{n} R_i^2 K_i / \sum_{i=1}^{n} K_i$. This variable, which represents the value-weighted $R^2$ on the market, has a mean of 5.25% over the entire sample period, with a median of 5.65% and an interquartile range of [3.68, 6.20]. Unexposed analyses show a diminishing trend of this measure, with values between 5% and 7% during the first years of the sample, and between 3% and 4% during the last years.

Table II provides descriptive statistics for the returns of UIRP, the market index assessed by the CRSP value-weighted index, and the Fama-French factor-mimicking portfolios. The latter are the "Small Minus Big" capitalization stocks (SMB), "High Minus Low" Book-to-Market stocks (HML), and "Momentum" (UMD for "Up Minus Down" previous returns) portfolios. The returns of these portfolios are extracted from the Kenneth French website[15]. Table II also presents the Pearson correlation coefficients between the returns of these portfolios.

(Insert Table II about here)

The average annualized return of UIRP on the sample period is 11.05%, which is lower than the average return of the market index, 11.80%. The standard deviation of UIRP, 18.28%, is higher than that of the market's index[16], 15.81%. UIRP exhibits a strong correlation with the market index, 0.977.

## 3 MUTUAL FUND PERFORMANCE ANALYSIS

This section starts by describing our mutual fund sample and proceeds to analyze mutual fund performance for both actively and passively managed equity mutual funds. It is important to recall that, as our analysis considers a framework with informational asymmetries between investors, the market portfolio is irrelevant as a reference portfolio for measuring fund performance.

### 3.1 DATA

We collect data for U.S. equity mutual funds from the CRSP Survivor-Bias-Free Mutual Fund Database. The analysis period is from January 1968 to December 2020, and our sample construction follows the same guidelines as in previous related empirical studies. We choose mutual funds that invest in domestic U. S. equities.

CRSP provides information on mutual funds' investment objectives from several sources: the CRSP "Strategic Insight Objective", which covers the period between 1993 and 1998; Wiesenberger, which covers the period from 1962 to 1993; and Lipper, which starts in 1998. The corresponding codes are synthesized by the CRSP "Style Code", which covers the entire database period and, therefore, will be used to construct our mutual fund sample. The CRSP mutual fund database guide provides a table of correspondence between the CRSP style code and the three funds' investment objectives mentioned above.

---

[15] http://mba.tuck.dartmouth.edu/pages/faculty/ken.french/data_library.html.
[16] The fact that the portfolio of uninformed investors is riskier than the market portfolio is quite natural. Informed investors tilt their portfolio toward securities with higher private information content. Because of the informational advantage, the risk of the portfolio held by informed investors is, *ceteris paribus*, lower than the risk of the portfolio held by uninformed investors. The market portfolio is the average of these two portfolios. It follows that the portfolio of uninformed investors is, *ceteris paribus*, riskier than the market portfolio.

Our investigations show that the CRSP style code is missing from some observations. At the same time, the fund's investment objective is still available at the same period in at least one of the other three classifications. We fill the missing data for such observations with the corresponding available code in the three other classifications. Moreover, suppose an investment objective code is available for a given fund during year $y-1$ but not during year $y$. In that case,en we replace the missing data during year $y$ with the data available during year $y-1$. This procedure is standard in existing empirical studies (e. g., Amihud & Goyenko, 2003).

The sample comprises funds for which the first level in the CRSP style classification is "Equity" (the corresponding letter in the database is "E"), with the second level being "Domestic" (the corresponding letter in the database is "D"). Domestic equity mutual funds are thus determined by the first two letters, "ED", of the CRSP style classification. At the third level, we retain three categories as follows: "Sector" (S), "Cap-based" (C), and "Style" (Y). At the fourth level, we choose the sub-categories "Large Cap" (L), "Mid Cap" (M), and "Small Cap" (S) within the "Cap-based" category, and the sub-categories "Growth" (G), "Growth & Income" (B) and "Income" (I) within the "Style" category. These categories contain a relatively similar number of funds. We exclude all other funds such as "foreign", "bond", "corporate" or "government" funds because they generally invest in minimal quantities of domestic U. S. equities. We exclude passively managed funds identified in the CRSP database by an index fund flag. We analyze these funds separately in our paper.

By applying all these restrictions, our final sample comprises 26,407 distinct actively managed U.S. domestic equity mutual funds that existed sometime during the 53 years from January 1968 to December 2020. Table III presents descriptive statistics for our sample funds.

(Insert Table III about here)

The majority of our sample funds are open-end funds. The average Total Net Assets (TNA) is 395.29 M$, the average expense ratio (expressed as the percentage of total net assets per year)

is 1.32%, and the average turnover ratio is 89.79% per year.

(Insert Figure I about here)

Figure I shows the value of one euro invested in UIRP, the CRSP value-weighted stock market index, an equally weighted portfolio of all sample funds, and a TNA-weighted portfolio of these funds, together with the SMB, HML, and UMD factor-mimicking portfolios. The two portfolios of mutual funds have a similar pattern to UIRP.

## 3.2 MUTUAL FUND PERFORMANCE

We measure fund performance by the alpha, the risk-adjusted return from the classical model proposed by Jensen (1968). This one-factor model traditionally considers the stock market index as the reference portfolio. According to the theory, we use our UIRP portfolio as a reference portfolio. We also employ performance models including factor-mimicking portfolios as proposed by Fama & French (1993) and Carhart (1997): SMB (small minus big size stocks); HML (high minus low book-to-market ratio stocks); and UMD (winner minus loser stocks). The general performance model specification is

$$r_{pt} - r_{ft} = Alpha + \sum_{j=1}^{n} \beta_j * r_{j,t} + \epsilon_t \qquad (12)$$

where $r_{pt}$ is the monthly net return of the portfolio of mutual funds over month $t$, $Alpha$ is the regression's intercept, $r_{ft}$ is the risk-free rate of return, $r_{j,t}$ is the return of the reference portfolio $j$ considered in the regression, $n$ is the number of reference portfolios used, and $\epsilon_t$ is the error term. In the particular case of the market portfolio, the return considered in this equation is in excess of the risk-free rate.

In what follows, we use Jensen's one-factor specification with the stock market index as a reference portfolio, and alternatively with UIRP, and Carhart's specification with the stock market index and alternatively with UIRP as reference portfolios. We estimate these regressions using OLS.

### 3.2.1 Overall performance

Rather than studying individual mutual funds, we focus on TNA-weighted portfolios of all sample funds. For a given month of a given year, the weight of a fund in the portfolio is the mutual fund's TNA as of the previous month. This procedure, standard in the existing literature, avoids eliminating mutual funds that do not survive during the sample period, for which there is insufficient data to determine their performance, thereby mitigating survivorship biases.

(Insert Table IV about here)

Table IV presents the alpha of the TNA-weighted portfolio of all mutual funds in the sample and the corresponding factor loadings (betas). With the one-factor model considering UIRP as the reference portfolio, mutual funds exhibit a positive but insignificant alpha of 0.23% per annum. The beta of the UIRP portfolio is 0.84. The portfolio of mutual funds displays a positive and insignificant alpha if the performance model includes the other factor-mimicking portfolios.

The results change dramatically if the stock market index is used as a reference portfolio. The one-factor model specification of Jensen (1968) shows a negative and highly significant alpha of -1.28%. With model specifications including factor-mimicking portfolios, alphas range between $[-1.31\%, -0.99\%]$. These results are in line with those found in the previous empirical literature[17] and show that, collectively, actively managed equity mutual funds underperform the market.

To conclude, if the performance model uses the appropriate reference portfolio, UIRP, then the average mutual fund's performance is positive and insignificant. The average mutual fund is negative and significant if the stock market index is used instead.

### 3.2.2 Analysis of different categories of funds

The degree of private information of the assets held by funds may vary with the fund's investment objective. For example, investing in small-capitalization stocks requires stronger expertise than for large-capitalization stocks because small firms are less well-known by the market. Table III shows that the average expense ratio for small-cap funds (1.45%) is higher than for large-cap funds (0.71%). According to the theory, the performance of the assets held by funds with stronger expertise is higher than for other funds, but the extra performance is compensated by the additional information acquisition costs incurred by the fund. It follows that the funds' alpha *net of expenses* should be indistinguishable from zero for all categories of funds, irrespective of their investment objective.

---

[17] For comparison with existing results, Wermers (2000) studies a sample of 1,788 actively managed equity mutual funds on the period between 1975 and 1994. The sample's alpha estimated with the Carhart's performance model is $-1.16\%$ per annum for the TNA-weighted portfolio of funds, and $-1.15\%$ for the equally weighted portfolio of funds, both significant at the 1% level. Kacperczyk et al. (2005) use a sample of 1,771 U.S. equity mutual funds on the period between 1984 and 1999. The Carhart's net alpha is a non-significant $-0.07\%$ per quarter for the equally weighted portfolio of sample mutual funds.

Table V presents the alpha of portfolios of funds by investment objective. The alphas based on UIRP are either insignificant or positive and weakly significant for all categories of funds. This result is in line with the theory. On the contrary, the alphas obtained with the stock market index (which is commonly used as a benchmark in the previous literature) are either insignificant or significantly negative. The market-based alpha is negative and significant for the categories Growth, Sector, and "Other". The results obtained with Carhart's model specification are relatively similar, whether one uses UIRP or the stock market index as a reference portfolio.

(Insert Table V about here)

Other characteristics that are known to be related to performance are fund expenses, turnover, and size. A puzzling fact about actively managed equity mutual funds resides in the strong negative relationship between performance and expenses. Gil-Bazo & Ruiz-VerdÚ (2009) explain this result by strategic fee setting when investors have different degrees of sensitivity to performance. According to our previous discussion, performance and expenses should be positively related if the performance is assessed with gross returns. A higher informational advantage requires higher information acquisition costs and generates higher performance. After subtracting expenses, the relationship between performance and expenses is expected to vanish.

(Insert Table VI about here)

Table VI, Panel A, presents the alpha of portfolios of funds by deciles of expense ratio. We rank mutual funds every month in ten groups based on their expense ratio during that month[18]. Then, we form TNA-value-weighted portfolios for each rank of expense ratio. We then calculate the alpha of these ten portfolios with several performance model specifications.

If we consider UIRP as a reference portfolio, the results show that mutual funds belonging to low- and medium-expense deciles exhibit either non-significant or significantly positive alphas. The portfolios corresponding to high-expense deciles exhibit a negative and significant alpha, especially for deciles 9 and 10. With the market-based portfolio, the alpha is negative for every expense decile and significant for all deciles except decile 1 and decile 2. The results obtained with the four-factor model specifications are similar. Overall, the results are similar to the previous literature, which shows that funds charging higher expenses tend to have lower performance if the latter is assessed using the stock market index as a reference portfolio. These results hold even if performance is assessed with UIRP.

---

[18] In some cases, expense ratios are missing. In that case, as is common in the literature, we replace the missing observation by the last available expense ratio, if any, for each mutual fund.

Turnover is another variable that is widely considered in studies of mutual fund performance, but there is no consensus in the literature about its impact on fund performance. This variable is likely to be related to performance because it determines operating costs and can indicate of informational advantages. Grinblatt & Titman (1994) find a positive relationship between fund performance and turnover and argue that this result is explained by the fact that fund managers trade more if they have more informational advantages. On the contrary, Amihud & Goyenko (2013) find that greater selectivity, indicative of superior informational advantages, is not associated with more frequent trading.

Table VI, Panel B, replicates the analysis by category of fund turnover. The procedure adopted for classifying mutual funds is the same as the one used for expenses. We rank mutual funds every month in ten groups based on their turnover ratio during that month. Then, we form TNA-value-weighted portfolios for each rank of turnover ratio and calculate the alpha of each of these ten portfolios with the same performance model specifications. The results show that the UIRP-adjusted alpha is insignificant for every category of fund turnover. Higher trading in mutual fund assets does not seem to be associated with higher performance. If the market portfolio is used instead, all categories of funds exhibit a negative alpha, which is significant for all deciles except deciles 1 and 9. Results are relatively similar by using four-factor specifications. Furthermore, there is no discernible pattern in the link between turnover and performance; this result is consistent with the existing literature.

The fund's size (TNA) is another well-known determinant of fund performance. The existing evidence shows, in general, that performance deteriorates if the fund becomes larger (diseconomies of scale) or older (e. g., Pástor *et al.*, 2015). Table VI, Panel C, presents fund alpha for deciles of fund total net assets built in the same way as for expenses and turnover. We rank mutual funds monthly in ten groups based on their TNA as of the previous month. Then, we form TNA-value-weighted portfolios for each TNA rank and calculate the alpha of each of these ten portfolios with the same performance model specifications. Irrespective of the funds' TNA category, the UIRP-adjusted alpha is insignificant for all portfolios, whether one uses the one- or four-factor specification. If the reference portfolio is the stock market index, then the alpha is negative and significant for most TNA deciles and performance model specifications. The negative and significant alphas obtained systematically with the market portfolio as the reference portfolio suggest that this portfolio is not efficient.

### 3.2.3 Performance and degree of active management (selectivity)

Using a relevant reference portfolio is not only important to assess overall fund performance; it is also crucial for studying the link between performance and the fund's degree of active management. The degree of active management is sometimes referred to as "selectivity". Stronger active management generates higher informational advantages and is indicative of investments in stocks with higher degrees of informational asymmetries. A higher selectivity moves the manager's optimal portfolio away from the market portfolio to a greater extent. Symmetrically, the reference portfolio of the uninformed investor will also deviate more from the market portfolio. It follows that the irrelevancy of the market portfolio as a reference portfolio for measuring performance increases as the fund's degree of active management is stronger.

The literature assesses the degree of active management for mutual funds in various ways. Kacperczyk *et al.* (2005) and Burlacu *et al.* (2006) consider the level of concentration of actively managed equity mutual fund portfolios in economic sectors and find that it is positively and significantly related to the fund's performance. Wermers (2003) uses the tracking error variance to measure fund selectivity and finds similar results. Cremers & Petajisto (2009) determine fund selectivity by the "active share", which measures the deviation of the mutual fund portfolios' weights from those of the market portfolio. Amihud & Goyenko (2013) measure fund selectivity by the fund's $R^2$ obtained with a multifactor model. The consensus in the empirical literature in the field is that funds that are more actively managed perform better.

We determine a fund's degree of active management with the mutual fund's $R^2$, a measure proposed by Amihud & Goyenko (2013). This measure is simple to calculate and obtains from a regression of fund returns on the returns of the stock market index and factor-mimicking portfolios. A higher $R^2$ indicates a stronger link with factor portfolios and hence is associated with lower selectivity (the fund tracks reference portfolios more closely).

We calculate $R^2$ for every fund with sufficient return data in our sample. (This $R^2$ is not to be confounded with the one we use for constructing UIRP.) Specifically, we decompose each fund's entire period of available data into three-year nonoverlapping periods. Then, we calculate a given fund's $R^2$ by projecting its returns on the corresponding factors for each period. We retain only funds and periods with a minimum number of 30 available monthly returns per period (among the 36 maximum monthly returns). $R^2$ represents the degree of fund active management during that period; the measure is updated each period for each fund.

As in Amihud & Goyenko (2013), we use two model specifications for determining $R^2$: a one-factor specification based on the stock market index and Carhart's four-factor specification. This procedure allows determining the $R^2$ for 20,526 mutual funds in our sample. With the classical CAPM specification, the mean $R^2$ is 0.78 with a median of 0.85 and an intra-quartile range of [0.72, 0.92]. With Carhart's specification, the mean $R^2$ is 0.86 with a median of 0.92 and an intra-quartile range of [0.85, 0.96].

In a second step, we rank $R^2$ in 10 deciles each month during the sample period, and then we build, for each decile, value-weighted portfolios of all mutual funds with available $R^2$. This generates *ten* portfolios of funds, ranking from the one that is the most "selective" (low $R^2$) to the one that is the most "diversified" (high $R^2$). The alpha of each portfolio is estimated on the entire period from 1968 to 2020 with the model specifications that are considered throughout our study. Table VII presents the alpha for each $R^2$ decile and each performance model.

(Insert Table VII about here)

Alpha is positive and insignificant for all UIRP-based (one- and four-factor) model specifications and most deciles, except for decile 6 with the one-factor specification. The alpha determined with model specifications that consider the market index as a reference portfolio is negative and significant in most cases. It is also noteworthy that mutual funds situated in deciles 1 and 2 display relatively higher alphas. As in previous studies, greater selectivity seems to generate higher alphas for funds with very high levels of specialization.

### 3.3 THE CASE OF INDEX MUTUAL FUNDS

We also separately analyze the case of U.S. domestic equity index mutual funds, which generally are considered "*passively*" managed portfolios. The investment objective of such funds is to track the market portfolio or specific market segments. Previous empirical studies exclude these types of funds from their performance analyses or use them as reference portfolios.

In the REE models' framework, tracking the market portfolio is not a passive type of management because it requires obtaining informational advantages. The market portfolio is not accessible to uninformed investors because it is unobservable. This portfolio is rather accessible to a representative investor possessing an "average" type of information, that is, which beliefs represent the weighted-average beliefs of all investors on the market (e.g., Biais *et al.*, 2010). This average type of information obtains at a cost. It follows that, as for actively managed equity mutual funds, the performance of "index" funds has to be judged in relationship with the reference portfolio from the perspective of the uninformed investor, not that of the index that managers track.

The CRSP mutual fund database classifies index funds into three categories. The first one comprises "index-based funds" (index flag = "B"), and utilizes indexes as "the primary filter for the purchase and sale of securities". The second one comprises "pure index funds" (index flag = "D"), which objective is to "match the total investment performance of a publicly recognized securities market index". The third category comprises "enhanced" index funds, (index flag = "E"), which objective is to "exceed the total investment performance of a publicly recognized securities market".

We replicate some of our analyses for 3,286 index funds identified by the CRSP survivorship-free mutual fund database. We build the sample of index funds by respecting the same criteria as for actively managed equity funds in our sample. Overall, index funds exhibit an expense ratio of 0.86%, a turnover ratio of 116%, and an average TNA of 1,164 M$. These figures suggest a high activity level, but this activity is mainly due to the presence of "enhanced" index funds. If the analysis is restricted only to "pure" index funds, the expense ratio equals 0.59%, the turnover ratio equals 60%, and the average TNA equals 2,012 M$.

As for actively managed funds, we find that the TNA-weighted portfolio of index funds exhibits an insignificant alpha of $-9$ basis points per year ($t$-value of $-0.12$) with the one-factor model based on our UIRP portfolio. With the four-factor model considering UIRP and factor-mimicking portfolios, we find a positive and insignificant alpha of $9$ basis points per year ($t$-value of $0.14$). Using the stock market index as a reference portfolio in the one-factor model, the alpha becomes negative and significant, with a value of $-145$ basis points per year ($t$-value of $-2.57$). The alpha from the Carhart's (1995) model is negative, $-113$ basis points per year, and significant ($t$-value of $-2.17$).

The previous literature suggests that investors should buy index funds because they perform better than actively managed funds. Our results show that if the performance is measured using the appropriate reference portfolio, UIRP, uninformed investors would be indifferent between index funds and actively managed funds because the alpha is insignificant for both categories. The negative and significant performance obtained by using the market portfolio as the reference portfolio indicates that this portfolio is not the correct benchmark.

## 3.4 ROBUSTNESS CHECKS

In unexposed analyses, we have checked the robustness of our results to various alternative model specifications and found similar results. These analyses are available at request. First, we have estimated each individual fund's alpha on the entire period of data availability with the various performance model specifications used in this article rather than constructing a portfolio of all sample mutual funds. Then, we have averaged these alphas across mutual funds using TNA-based weights (and equal weights) to analyze fund performance. Our results do not change fundamentally. However, in many cases, we find positive and significant alphas using UIRP. Overall, these results provide a more optimistic picture of mutual fund active management. The problem with this empirical approach used in some existing studies is that it may generate a survivorship bias. Indeed, this analysis implicitly considers only funds with sufficient data to estimate performance accurately. Most funds with enough data are funds that survive and are known to exhibit higher performance. Because this type of analysis implicitly considers funds that perform better, the alphas are improved, which is what we observe empirically.

We have also replicated our analyses using equally-weighted portfolios rather than TNA-weighted ones. Again, our results do not change fundamentally. Still, they show a more optimistic picture of active mutual fund management in that the alphas calculated with UIRP are higher than those calculated with the methodology used in the paper. The better results obtained with equally weighted portfolios may be explained by the fact that this methodology gives more importance to small-TNA funds, which tend to exhibit higher performance, as suggested by our results in Table VI (diseconomies of scale). We also find that the alpha adjusted for market risk is negative and significant with most performance model specifications and for most categories of funds.

# CONCLUSION

Our paper proposes a new approach for assessing the performance of actively managed equity mutual funds. While the related literature stresses some doubt about the expertise of the funds' managers and their ability to create value for their clients, one can still observe a high demand of individual investors for such investment vehicles. The objective of this paper is to bring an explanation for this by mobilizing the theory of portfolio choice under asymmetric information. Active management is justified by the existence of informational asymmetries between mutual fund managers (considered informed investors) and their clients (considered uninformed investors). When investors do not share homogeneous information about the future payoff of risky assets, the traditional CAPM does not hold anymore. The theory of rational expectation equilibrium models (in particular, the GS model) teaches us that when there are informed and uninformed investors on the market, none of them hold the traditional market portfolio. Still, both must consider a portfolio depending on their corresponding available information set.

In this paper, we build on this literature and propose a new reference portfolio, as an alternative to the traditional market portfolio, for assessing mutual fund performance from their clients' perspective, considered uninformed investors. Based on the GS model, we first show that the uninformed investor's reference portfolio tilts towards stocks with low levels of information asymmetry (high levels of information precision). The first contribution of this paper is to propose a theoretical approach to building the uninformed investor's (i.e., the mutual funds' client) reference portfolio. Then we mobilize this approach to construct this portfolio empirically. To achieve this goal, we use the "Proxy E(R)" measure of Burlacu et al. (2002) and compute its value for each stock and each month of our sample period. The second contribution is to empirically analyze the performance of actively managed U.S. domestic equity mutual funds by using the uninformed investor's reference portfolio in Jensen's (1968) and Carhart's (1997) performance model specifications. Following GS, the alpha, computed with returns net of expenses, should be insignificantly different from zero from the perspective of the uninformed investor on an informationally efficient market, as expenses exactly compensate for the information acquisition costs. We precisely obtain these results.

By using the correct reference portfolio from the uninformed investor's perspective (UIRP), we rehabilitate to some extent the value of active management: managers of equity mutual funds seem to provide non-negative UIRP-adjusted alphas to their clients. This may explain why uninformed economic agents continue to use mutual funds services and why investment in mutual funds is so important, despite the pessimistic results obtained by a body of the existing empirical literature. To some extent, our results are in line with those obtained by Wermers (2000): fund managers perform well enough to compensate for expenses and transaction costs.

This study may open some new avenues for research. The literature offers several measures for assessing the degree of information asymmetry between investors in the market. Such measures may prove fruitful in constructing the uninformed investor's reference portfolio in alternative ways to that offered in our paper. The UIRP-based performance approach may also provide answers to the challenging debate on "luck versus skill" in cross-sectional studies of mutual fund performance. Fama & French (2010) put forward the "arithmetic of active management" (Sharpe, 1991), according to which active management is a zero-sum game. They find that few actively managed funds produce risk-adjusted returns that cover their costs. However, the results obtained by the existing literature are far from offering a consensus on this debate (e.g., Kosowski *et al.*, 2006).

Research on the persistence of mutual fund performance, which does not offer a consensus, could also benefit from using the correct reference portfolio. The seminal study of Carhart (1997) shows that common factors almost completely explain persistence in stock returns. Similarly, Fama & French (2010) and Berk & van Binsbergen (2015) find little evidence of persistence in actively managed equity mutual funds. However, Grinblatt & Titman (1992) find that differences in performance between funds persist over time and argue that this persistence shows the ability of fund managers to generate abnormal returns. More recent studies corroborate this result at the international level (e. g., Ferreira *et al.*, 2019). All these studies have in common that they use the stock market index, among others, as the reference portfolio. We believe that the UIRP-based approach may provide additional insights into these issues. Existing studies suggest that fund performance is highly sensitive to reference portfolio specifications (Grinblatt & Titman, 1994).

# TABLES AND FIGURES

## Table I
### Simulation Analysis

This table considers a GS setting with two risky assets. It presents the relationship between noise (supply uncertainty $z_{11}$ for stock 1) and three related variables: the relative degree of information asymmetry for asset 1, the relative weight of asset 1 in the price-contingent portfolio assessed with unobservable REE parameters, and the relative weight of asset 1 in the price-contingent portfolio assessed with the observable variable $1-$"$Proxy\ E(r)$". The degree of information asymmetry of asset 1 relative to that of asset 2 is measured by $\frac{n_{11}}{n_{22}}$, where $n_{11}$ and $n_{22}$ are the diagonal terms of the matrix $N$. The row heading for this variable is "Information Asymmetry". The weight of asset 1 relative to asset 2 in the price-contingent portfolio with unobservable REE parameters is measured by $\frac{E(\tilde{X}_{UI}^1)}{E(\tilde{X}_{UI}^2)}$, where $E(\tilde{X}_{UI}^1)$ and $E(\tilde{X}_{UI}^2)$ are the number of assets in the portfolio as given by Eq. (6). The corresponding row heading for this variable is "Asset Weight in Portfolio 1". The relative weight of asset 1 in the price-contingent portfolio based on the observable variable $1-$"$Proxy\ E(r)$" is $\frac{[1-"Proxy\ E(r_1)"]*E(\tilde{z}_1)}{[1-"Proxy\ E(r_2)"]*E(\tilde{z}_2)}$, where $E(\tilde{z}_i)$ is the expected supply for asset $i$, and $1-$"$Proxy\ E(r_i)$" is the variable $1-$"$Proxy\ E(r)$" determined for asset $i$. The corresponding row heading for this variable is "Asset weight in portfolio 2". In all simulations, the proportion of informed investors is 40%, and the risk-aversion coefficient is 0.1.

| Noise for stock 1 ($z_{11}$) | 1 | 2 | 3 | 4 | 5 | 6 | 7 | 8 | 9 | 10 |
|---|---|---|---|---|---|---|---|---|---|---|
| No correlation between securities : $t_{12}=u_{12}=z_{12}=0$ | | | | | | | | | | |
| Information Asymmetry | 1.00 | 1.89 | 2.68 | 3.40 | 4.05 | 4.64 | 5.17 | 5.67 | 6.12 | 6.54 |
| Asset Weight in Portfolio 1 | 1.00 | 0.98 | 0.96 | 0.95 | 0.93 | 0.92 | 0.91 | 0.90 | 0.89 | 0.89 |
| Asset Weight in Portfolio 2 | 1.00 | 0.95 | 0.91 | 0.88 | 0.85 | 0.82 | 0.79 | 0.77 | 0.75 | 0.74 |
| Correlation between information uncertainties : $t_{12}=0.5, u_{12}=z_{12}=0$ | | | | | | | | | | |
| Information asymmetry | 1.00 | 1.48 | 1.81 | 2.05 | 2.24 | 2.38 | 2.50 | 2.60 | 2.68 | 2.75 |
| Asset weight in portfolio 1 | 1.00 | 0.97 | 0.95 | 0.92 | 0.91 | 0.89 | 0.88 | 0.86 | 0.85 | 0.84 |
| Asset weight in portfolio 2 | 1.00 | 0.96 | 0.94 | 0.91 | 0.89 | 0.87 | 0.85 | 0.84 | 0.82 | 0.81 |
| Correlation between residual uncertainties : $u_{12}=0.5, t_{12}=z_{12}=0$ | | | | | | | | | | |
| Information asymmetry | 1.00 | 1.86 | 2.60 | 3.25 | 3.82 | 4.33 | 4.79 | 5.20 | 5.57 | 5.91 |
| Asset weight in portfolio 1 | 1.00 | 0.98 | 0.96 | 0.95 | 0.94 | 0.93 | 0.92 | 0.91 | 0.91 | 0.90 |
| Asset weight in portfolio 2 | 1.00 | 0.95 | 0.91 | 0.88 | 0.85 | 0.83 | 0.81 | 0.79 | 0.77 | 0.75 |
| Correlation between supply uncertainties : $z_{12}=0.5, t_{12}=u_{12}=0$ | | | | | | | | | | |
| Information asymmetry | 1.00 | 1.90 | 2.71 | 3.43 | 4.08 | 4.68 | 5.22 | 5.72 | 6.17 | 6.59 |
| Asset weight in portfolio 1 | 1.00 | 0.98 | 0.96 | 0.95 | 0.93 | 0.92 | 0.91 | 0.90 | 0.89 | 0.89 |
| Asset weight in portfolio 2 | 1.00 | 0.95 | 0.91 | 0.88 | 0.85 | 0.82 | 0.79 | 0.77 | 0.75 | 0.74 |

## Table II

**Portfolios' Summary Statistics and Pearson Correlation Coefficients**

This table presents percentage average annual returns, annual return standard deviation, and Pearson correlation coefficients for the monthly returns of our "uninformed-investor" reference portfolio (UIRP), the market portfolio proxy considered as the CRSP value-weighted stock index, and three factor-mimicking portfolios. The UIRP portfolio over (under) weighs stocks with low (high) levels of private information content. The level of private information content is measured by "Proxy E(r)", a measure proposed by Burlacu *et al.* (2012). The factor-mimicking portfolios are the "SMB" (for "Small Minus Big" capitalization stocks), "HML" (for "High Minus Low" book-to-market stocks), and "UMD" (for "Up Minus Down" prior stock returns) portfolios. The analysis period is from January 1968 to December 2020. Three, two, and one star indicate a significance level of 0.01, 0.05, and 0.10, respectively.

| Portfolio | Mean Return (%/year) | Standard Deviation of Return (%/year) | Pearson Correlation Coefficients between Returns | | | |
|---|---|---|---|---|---|---|
| | | | UIRP | Stock market index | SMB | HML |
| UIRP | 11.05 | 18.28 | | | | |
| Stock market index | 11.80 | 15.81 | 0.977*** | | | |
| SMB | 1.67 | 10.64 | 0.406*** | 0.298*** | | |
| HML | 3.07 | 10.19 | -0.277*** | -0.225*** | -0.174*** | |
| UMD | 7.53 | 14.95 | -0.178** | -0.166*** | -0.058*** | -0.201** |

## Table III
**Summary Statistics for Mutual Funds**

This table presents fund characteristics for 26,407 U.S. actively managed equity mutual funds on the period from January 1968 to December 2020. "TNA" (Total Net Assets) is the closing market value of securities owned, plus all assets minus all liabilities. "Turnover" (over the calendar year) is the minimum of aggregate purchases of securities or aggregate sales of securities divided by the average TNA of the fund. "Expenses" (over the calendar year) designates the percentage of the total investment shareholders pay for the mutual fund's operating expenses. Net fund returns are calculated as follows. Each month of each year, we compute the TNA-weighted mean of net returns for all sample funds, where the weights are the TNA at the end of the previous month. Then we average returns across time during the whole period with equal weights. "NA" means that the information is not available in the database.

| Group of funds | | Number of funds | Net returns (%/year) | TNA ($M) | Expense Ratio (%/year) | Turnover Rate (%/year) |
|---|---|---|---|---|---|---|
| All funds | | 26,407 | 10.31 | 395.29 | 1.32 | 89.79 |
| Investment objective | Growth | 8,861 | 10.59 | 254.76 | 1.38 | 100.29 |
| | Growth & Income | 5,772 | 11.17 | 426.87 | 0.95 | 56.26 |
| | Income | 951 | 11.03 | 347.09 | 1.27 | 72.56 |
| | Large-Cap | 310 | 11.15 | 1,990.92 | 0.71 | 16.17 |
| | Mid-Cap | 2,280 | 14.66 | 246.26 | 1.40 | 103.18 |
| | Small-Cap | 3,416 | 10.10 | 188.35 | 1.45 | 99.96 |
| | Sector | 3,226 | 9.99 | 167.96 | 1.56 | 130.42 |
| | Other | 1,591 | 3.63 | 107.95 | 1.70 | 288.14 |
| Open fund | Yes | 12,116 | 11.48 | 637.27 | 1.27 | 130.82 |
| | No | 91 | 10.63 | 290.01 | 0.95 | 45.07 |
| | NA | 14,200 | 10.33 | 308.97 | 1.31 | 81.88 |
| Retail fund | Yes | 11,070 | 10.63 | 317.03 | 1.34 | 88.26 |
| | No | 12,934 | 11.32 | 268.92 | 1.01 | 66.55 |
| | NA | 2,403 | 7.94 | 190.07 | 1.29 | 92.82 |
| Institutional fund | Yes | 8,143 | 11.41 | 289.29 | 0.90 | 67.72 |
| | No | 15,861 | 10.54 | 296.39 | 1.32 | 86.00 |
| | NA | 2,403 | 7.94 | 190.07 | 1.29 | 92.82 |
| Dead fund | Yes | 12,364 | 9.33 | 143.77 | 1.37 | 88.19 |
| | No | 12,831 | 10.77 | 379.45 | 1.15 | 81.52 |
| | NA | 1,212 | 11.09 | 335.24 | 0.77 | 64.09 |

## Table IV
**Performance models' estimates**

This table presents coefficient estimates of the regression of monthly returns of the value-weighted portfolio of mutual funds on the returns of reference portfolios on the period between January 1968 and December 2020. The portfolio of funds comprises 26,407 U.S. domestic actively managed equity funds that existed sometime during this period. The set of reference portfolios comprises UIRP, the CRSP value-weighted index considered as the proxy for the market portfolio, the SMB (for "Small Minus Big" capitalization stocks), "HML" (for "High Minus Low" book-to-market stocks), and "UMD" (for "Up Minus Down" prior stock returns) factor-mimicking portfolios. Alpha is estimated using various portfolios in combination. The general performance model specification is

$$r_{pt} - r_{ft} = Alpha + \sum_{j=1}^{n} \beta_j * r_{j,t} + \epsilon_t$$

where $r_{pt}$ is the monthly net return of the portfolio of funds over month $t$, $Alpha$ is the regression's intercept, $r_{ft}$ is the risk-free rate of return, $r_{j,t}$ is the return of the reference portfolio considered in the regression, $n$ is the number of reference portfolios used, and $\epsilon_t$ is the error term. The return of the stock index portfolio is considered in excess of the risk-free rate. The coefficient's *t*-statistics are provided between parentheses. Three, two, and one star indicate a significance level of 0.01, 0.05, and 0.10, respectively.

| MODEL | ALPHA (%/year) | Beta UIRP | Beta Market | Beta SMB | Beta HML | Beta UMD | $R^2$ |
|---|---|---|---|---|---|---|---|
| Models with UIRP as a reference portfolio | 0.23 (0.48) | 0.84 (117.21)*** | - | - | - | - | 0.956 |
| | 0.23 (0.51) | 0.85 (109.66)*** | | -0.05 (-3.82)*** | | | 0.957 |
| | 0.31 (0.67) | 0.85 (96.807)*** | | -0.05 (-3.92)*** | -0.02 (-1.47) | | 0.957 |
| | 0.00 (0.01) | 0.86 (105.06)*** | - | -0.05 (-3.96)*** | -0.01 (-0.54) | 0.03 (3.36)*** | 0.958 |
| Models with the stock market index as a reference portfolio | -1.28 (-3.24)*** | - | 0.98 (135.58)*** | - | - | - | 0.967 |
| | -1.31 (-3.65)*** | - | 0.95 (139.23)*** | 0.12 (11.87)*** | | | 0.973 |
| | -0.99 (-2.87)** | - | 0.94 (142.17)*** | 0.11 (11.42)*** | -0.08 (-8.16)*** | | 0.975 |
| | -1.06 (-3.01)*** | - | 0.95 (139.18)*** | 0.11 (11.45)*** | -0.08 (-7.66)*** | 0.01 (0.97) | 0.975 |

## Table V
### Performance models' estimates by fund self-reported investment objective

This table presents coefficient estimates for the regression of monthly returns of value-weighted portfolios of mutual funds on the returns of reference portfolios on the period between January 1968 and December 2020. The sample comprises 26,407 U.S. domestic actively managed equity funds that existed sometime during this period. For each month during this period, monthly returns are averaged with the previous month's TNA weights across all funds having the same investment objective during that month. There are eight mutual fund portfolios corresponding to funds' investment objectives, as indicated in the table. The set of reference portfolios comprises UIRP, the CRSP value-weighted index considered as a proxy of the market portfolio, the SMB (for "Small Minus Big" capitalization stocks), "HML" (for "High Minus Low" book-to-market stocks), and "UMD" (for "Up Minus Down" prior stock returns) factor-mimicking portfolios. Alpha is estimated using various reference portfolios in combination. The general performance model specification is

$$r_{pt} - r_{ft} = Alpha + \sum_{j=1}^{n} \beta_j * r_{j,t} + \epsilon_t$$

where $r_{pt}$ is the monthly net return of the portfolio of funds over month $t$, $Alpha$ is the regression's intercept, $r_{ft}$ is the risk-free rate of return, $r_{j,t}$ is the return of the reference portfolio considered in the regression, $n$ is the number of reference portfolios used, and $\epsilon_t$ is the error term. The return of the stock index portfolio is considered in excess of the risk-free rate. The coefficient's *t*-statistics are provided between parentheses. Three, two, and one star indicate a significance level of 0.01, 0.05, and 0.10, respectively.

| Factors included in performance models | Investment objective | | | | | | | |
|---|---|---|---|---|---|---|---|---|
| | Growth | Growth & Income | Income | Large-Cap | Mid-Cap | Small-Cap | Sector | Other |
| UIRP | 0.433 | 1.62 | 1.669 | 1.474 | 2.206 | 0.373 | -1.88 | -1.039 |
| | (0.935) | (1.589) | (1.66)* | (1.669)* | (2.098)** | (0.329) | (-1.142) | (-1.079) |
| Market | -1.098 | 0.12 | -0.407 | -0.598 | 0.012 | -0.896 | -3.553 | -1.844 |
| | (-2.681)*** | (0.138) | (-0.57) | (-1.13) | (0.008) | (-0.765) | (-1.997)** | (-2.097)** |
| UIRP, SMB, HML, UMD | 0.264 | 1.644 | 0.553 | 0.638 | 1.742 | 0.18 | -2.76 | -1.265 |
| | (0.598) | (1.656)* | (0.776) | (1.014) | (1.842)* | (0.153) | (-1.898)* | (-1.467) |
| Market, SMB, HML, UMD | -0.801 | 0.565 | -0.694 | -0.563 | 0.276 | -0.658 | -3.981 | -1.844 |
| | (-2.318)** | (0.613) | (-1.318) | (-1.349) | (0.321) | (-0.574) | (-2.831)*** | (-2.263)** |

## Table VI
### Alphas by deciles of expenses, turnover, and total net assets

This table presents the intercept (alpha) from the regression of monthly returns of value-weighted portfolios of mutual funds on the returns of reference portfolios in the period between January 1968 and December 2020. The sample comprises 26,407 U.S. domestic actively managed equity funds that existed sometime during this period. For each month during this period, monthly returns are averaged based on the previous month's TNA weights across all funds belonging to the same decile of fund characteristics during that month. The characteristics considered are expenses, turnover, and TNA (Total Net Assets). TNA is the closing market value of securities owned, plus all assets minus all liabilities. "Turnover" (over the calendar year) is the minimum of aggregate purchases of securities or aggregate sales of securities divided by the average TNA of the fund. "Expenses" (over the calendar year) designates the percentage of the total investment that shareholders pay for the mutual fund operating expenses. Each month, funds are ranked in ten deciles according to the corresponding characteristic. Then, each month we form a TNA-weighted portfolio of all funds during a given decile, which gives us ten portfolios of mutual funds that correspond to the ten deciles considered. The set of reference portfolios comprises UIRP, the CRSP value-weighted index considered as a proxy for the market portfolio, the SMB (for "Small Minus Big" capitalization stocks), "HML" (for "High Minus Low" book-to-market stocks), and "UMD" (for "Up Minus Down" prior stock returns) factor-mimicking portfolios. Alpha is estimated using various reference portfolios in combination. The general performance model specification is

$$r_{pt} - r_{ft} = Alpha + \sum_{j=1}^{n} \beta_j * r_{j,t} + \epsilon_t$$

where $r_{pt}$ is the monthly net return of the portfolio of funds over month $t$, $Alpha$ is the regression's intercept, $r_{ft}$ is the risk-free rate of return, $r_{j,t}$ is the return of the reference portfolio portfolio considered in the regression, $n$ is the number of reference portfolios used, and $\epsilon_t$ is the error term. The return of the stock index portfolio is considered in excess of the risk-free rate. The coefficient's *t*-statistics are between parentheses. Three, two, and one star indicate a significance level of 0.01, 0.05, and 0.10, respectively.

Panel A: Fund alphas (%/year) by deciles of fund expenses.

| Factors included in performance models | Decile 1 (Low Expenses) | Decile 2 | Decile 3 | Decile 4 | Decile 5 | Decile 6 | Decile 7 | Decile 8 | Decile 9 | Decile 10 (High Expenses) |
|---|---|---|---|---|---|---|---|---|---|---|
| UIRP | 1.158 (1.647) | 1.401 (2.016)** | 0.373 (0.726) | -0.192 (-0.375) | -0.156 (-0.272) | -0.67 (-1.184) | -1.075 (-1.867)* | -1.229 (-2.034)** | -1.679 (-2.673)*** | -2.959 (-3.654)*** |
| Market | -0.216 (-0.411) | -0.156 (-0.243) | -1.087 (-2.187)** | -1.667 (-3.27)*** | -1.667 (-2.719)*** | -2.186 (-3.488)*** | -2.55 (-4.076)*** | -2.62 (-3.722)*** | -3.052 (-4.108)*** | -4.189 (-4.146)*** |
| UIRP, SMB, HML, UMD | 1.292 (1.897)* | 1.352 (1.909)* | -0.012 (-0.034) | -0.61 (-1.215) | -0.872 (-1.583) | -1.324 (-2.396)** | -1.62 (-2.836)*** | -1.584 (-2.681)*** | -2.139 (-3.424)*** | -3.309 (-4.552)*** |
| Market, SMB, HML, UMD | 0.24 (0.463) | 0.252 (0.398) | -1.039 (-2.287)** | -1.643 (-4.016)*** | -1.927 (-3.891)*** | -2.374 (-4.773)*** | -2.632 (-4.937)*** | -2.526 (-4.316)*** | -3.041 (-4.639)*** | -4.051 (-4.981)*** |

Panel B: Fund alphas (%/year) by deciles of fund turnover.

| Factors included in performance models | Decile 1 (Low Turnover) | Decile 2 | Decile 3 | Decile 4 | Decile 5 | Decile 6 | Decile 7 | Decile 8 | Decile 9 | Decile 10 (High Turnover) |
|---|---|---|---|---|---|---|---|---|---|---|
| UIRP | 0.674 (0.917) | 0.445 (0.795) | 0.373 (0.709) | 0.324 (0.57) | 0.529 (0.967) | -0.264 (-0.45) | -0.156 (-0.265) | -0.276 (-0.448) | 0.433 (0.429) | -0.956 (-1.296) |
| Market | -0.813 (-1.296) | -1.075 (-2.39)** | -1.146 (-2.697)*** | -1.205 (-2.505)** | -0.991 (-1.961)* | -1.809 (-3.351)*** | -1.691 (-2.639)*** | -1.809 (-2.69)*** | -1.087 (-1.035) | -2.268 (-2.584)*** |
| UIRP, SMB, HML, UMD | 0.553 (0.76) | 0.529 (0.928) | -0.18 (-0.361) | 0.312 (0.558) | 0.337 (0.603) | -0.598 (-1.018) | -0.61 (-1.121) | -0.98 (-1.708)* | 0.108 (0.101) | -2.08 (-3.362)*** |
| Market, SMB, HML, UMD | -0.503 (-0.795) | -0.574 (-1.31) | -1.253 (-3.004)*** | -0.765 (-1.675)* | -0.73 (-1.563) | -1.691 (-3.503)*** | -1.667 (-3.226)*** | -2.033 (-3.699)*** | -0.932 (-0.933) | -2.912 (-4.308)*** |

Panel C: Fund alphas (%/year) by deciles of fund total net assets.

| Factors included in performance models | Decile 1 (Low Total Net Assets) | Decile 2 | Decile 3 | Decile 4 | Decile 5 | Decile 6 | Decile 7 | Decile 8 | Decile 9 | Decile 10 (High Total Net Assets) |
|---|---|---|---|---|---|---|---|---|---|---|
| UIRP | 1.049 (1.013) | 0.602 (0.928) | 0.276 (0.498) | 0.409 (0.783) | 0.638 (1.302) | 0.445 (0.89) | 0.457 (0.913) | -0.096 (-0.189) | -0.144 (-0.283) | 0.337 (0.661) |
| Market | -0.036 (-0.034) | -0.658 (-0.963) | -1.063 (-1.846)* | -0.98 (-1.782)* | -0.801 (-1.502) | -1.027 (-2.041)** | -1.027 (-2.06)** | -1.608 (-3.102)*** | -1.643 (-3.252)*** | -1.193 (-2.838)*** |
| UIRP, SMB, HML, UMD | 1.037 (0.992) | 0.337 (0.508) | -0.024 (-0.049) | -0.048 (-0.093) | 0.048 (0.092) | -0.132 (-0.27) | -0.216 (-0.427) | -0.586 (-1.115) | -0.658 (-1.266) | 0.252 (0.499) |
| Market, SMB, HML, UMD | 0.276 (0.273) | -0.527 (-0.834) | -0.956 (-1.946)* | -1.015 (-2.202)** | -0.956 (-2.252)** | -1.17 (-2.897)*** | -1.253 (-3.073)*** | -1.655 (-4.013)*** | -1.726 (-4.084)*** | -0.825 (-2.101)** |

## Table VII

### Alphas by deciles of fund degree of active management, $R^2$

This table presents the intercept (alpha) from the regression of monthly returns of value-weighted portfolios of mutual funds on the returns of reference portfolios on the period between January 1968 and December 2020. The sample comprises 20,526 U.S. domestic actively managed equity funds that existed sometime during this period and for which there is sufficient data to calculate the degree of active management. For each month during this period, monthly returns are averaged using the previous month's TNA weights across all funds belonging to the same decile of fund degree of active management during that month. We measure active management as follows. We decompose the period of fund data availability in three-year periods and then calculate a given fund's $R^2$ by projecting its monthly returns on the returns of the stock market index (market model) or the stock market index together with the factor-mimicking portfolios used in Carhart (1997). We retain only funds with a minimum number of 30 available monthly returns per period. $R^2$ represents the degree of fund active management for the period during which it is determined and is updated each period. The portfolio of funds in each decile is determined as follows. First, each month funds are ranked in ten deciles according to their $R^2$. Then, we form a TNA-weighted portfolio of all funds in a given decile, which gives us ten portfolios of mutual funds that correspond to the ten deciles considered. The set of reference portfolios comprises UIRP, the CRSP value-weighted index considered as a proxy of the market portfolio, the SMB (for "Small Minus Big" capitalization stocks), "HML" (for "High Minus Low" book-to-market stocks), and "UMD" (for "Up Minus Down" prior stock returns) factor-mimicking portfolios. Alpha is estimated using various reference portfolios in combination. The general performance model specification is

$$r_{pt} - r_{ft} = Alpha + \sum_{j=1}^{n} \beta_j * r_{j,t} + \epsilon_t$$

where $r_{pt}$ is the monthly net return of the portfolio of funds over month $t$, $Alpha$ is the regression's intercept, $r_{ft}$ is the risk-free rate of return, $r_{j,t}$ is the return of the reference portfolios considered in the regression, $n$ is the number of reference portfolios used, and $\epsilon_t$ is the error term. The return of the stock index portfolio is considered in excess of the risk-free rate. The coefficient's *t*-statistics are between parentheses. Three, two, and one star indicate a significance level of 0.01, 0.05, and 0.10, respectively.

Panel A: Fund alphas (%/year) by deciles of fund degree of active management ($R^2$). $R^2$ is estimated with the market model.

| Factors included in performance models | Decile 1 (Low $R^2$=high degree of active management) | Decile 2 | Decile 3 | Decile 4 | Decile 5 | Decile 6 | Decile 7 | Decile 8 | Decile 9 | Decile 1 (High $R^2$=low degree of active management) |
|---|---|---|---|---|---|---|---|---|---|---|
| UIRP | 2.084 (1.423) | 1.037 (0.863) | -0.395 (-0.485) | 0.626 (0.918) | 0 (-0.004) | -0.991 (-1.718)* | -0.204 (-0.403) | 0.747 (1.57) | 0.168 (0.338) | 0.361 (0.652) |
| Market | 1.11 (0.769) | -0.359 (-0.301) | -1.903 (-2.368)** | -0.872 (-1.225) | -1.549 (-2.304)** | -2.55 (-4.005)*** | -1.773 (-3.342)*** | -0.825 (-1.799)* | -1.383 (-3.87)*** | -1.276 (-4.525)*** |
| UIRP, SMB, HML, UMD | 1.17 (0.788) | 0.517 (0.422) | -0.991 (-1.221) | -0.371 (-0.596) | -0.479 (-0.846) | -0.825 (-1.45) | -0.216 (-0.452) | 0.433 (0.954) | -0.012 (-0.021) | 0.084 (0.174) |
| Market, SMB, HML, UMD | 0.469 (0.322) | -0.515 (-0.441) | -2.092 (-2.928)*** | -1.442 (-2.63)*** | -1.561 (-3.039)*** | -1.915 (-3.739)*** | -1.3 (-2.957)*** | -0.646 (-1.629) | -1.11 (-3.163)*** | -1.087 (-3.869)*** |

Panel A: Fund alphas (%/year) by deciles of fund degree of active management ($R^2$). $R^2$ is estimated with Carhart's model.

| Factors included in performance models | Decile 1 (Low $R^2$=high degree of active management) | Decile 2 | Decile 3 | Decile 4 | Decile 5 | Decile 6 | Decile 7 | Decile 8 | Decile 9 | Decile 1 (High $R^2$=low degree of active management) |
|---|---|---|---|---|---|---|---|---|---|---|
| UIRP | 2.071 (1.405) | 1.243 (1.069) | 0.12 (0.182) | 0.614 (1.151) | -0.264 (-0.452) | -0.072 (-0.169) | -0.204 (-0.382) | 0.012 (0.016) | -0.048 (-0.094) | 0.578 (1.024) |
| Market | 1.122 (0.769) | -0.12 (-0.101) | -1.348 (-2.064)** | -0.884 (-1.586) | -1.797 (-2.945)*** | -1.584 (-3.105)*** | -1.75 (-3.045)*** | -1.561 (-3.337)*** | -1.679 (-4.176)*** | -1.098 (-3.441)*** |
| UIRP, SMB, HML, UMD | 1.085 (0.721) | 0.916 (0.766) | -0.658 (-0.962) | 0.132 (0.244) | -0.551 (-0.958) | -0.479 (-1.127) | -0.622 (-1.212) | -0.108 (-0.211) | -0.108 (-0.2) | 0.264 (0.49) |
| Market, SMB, HML, UMD | 0.421 (0.279) | -0.048 (-0.046) | -1.703 (-2.718)*** | -0.896 (-1.804)* | -1.608 (-3.033)*** | -1.502 (-3.625)*** | -1.679 (-3.608)*** | -1.205 (-3.036)*** | -1.276 (-3.391)*** | -0.956 (-2.983)*** |

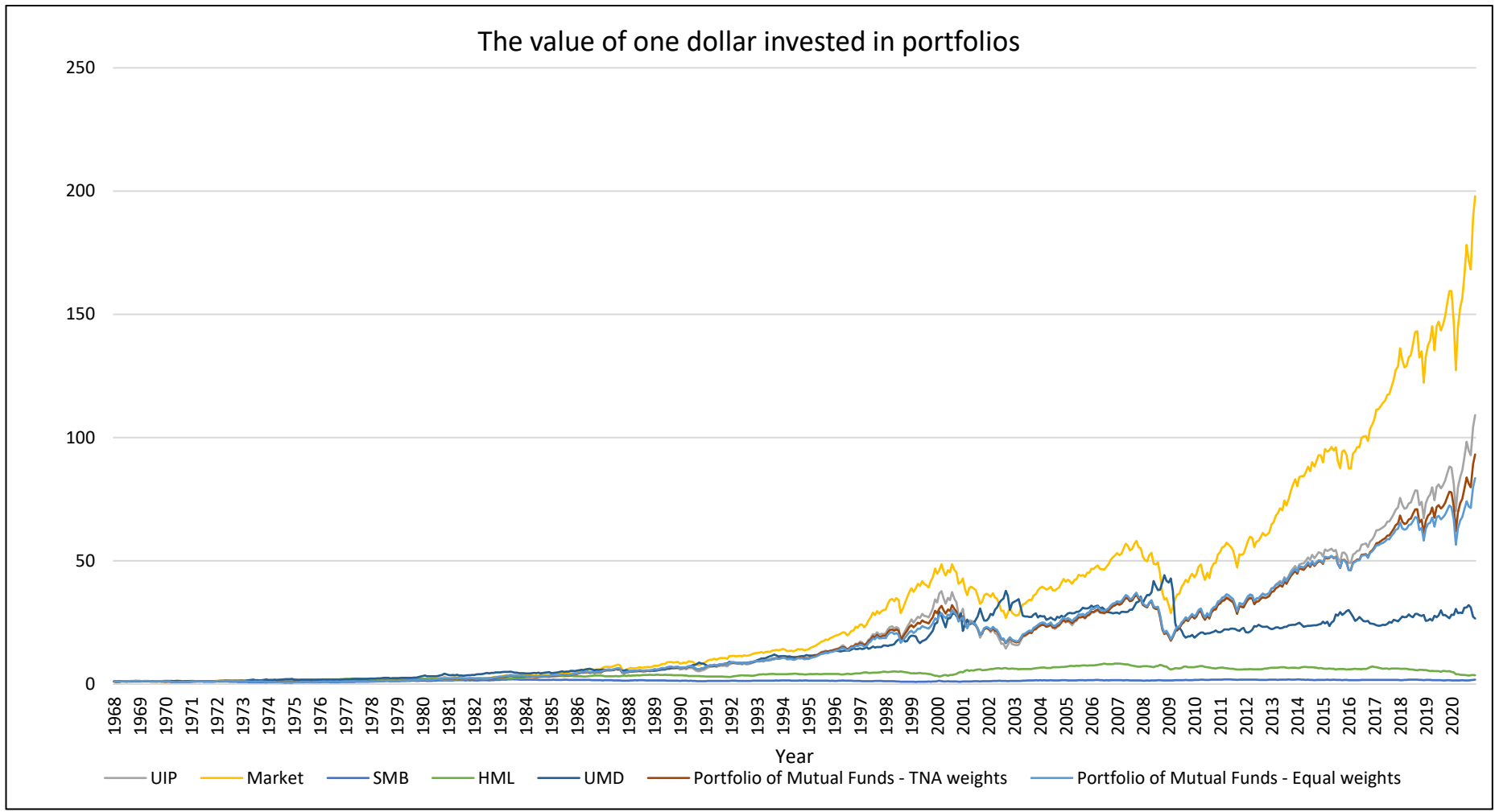

**Figure I.** This figure presents the value of one dollar invested in reference portfolio portfolios (with dividends reinvested). The portfolios are our uninformed investors' reference portfolio (UIRP), the CRSP value-weighted stock market index, which proxies for the market portfolio, the Fama-French SMB, HML, and UMD factor-mimicking portfolios, together with an equally-weighted and a TNA-weighted portfolio of 26,407 U.S. domestic actively managed equity mutual funds that existed sometime during the period from January 1968 to December 2020. UIRP under-weighs (overweighs) stocks with high (low) degrees of informational asymmetries. The analysis period is from January 1968 to December 2020.